\documentclass[twocolumn,envcountsame]{svjour3}

\newif\iflong
\longtrue

\usepackage[T1]{fontenc} 
\usepackage[utf8]{inputenc}
\usepackage{times}

\usepackage[pdftex]{graphicx}
\usepackage{amsmath,amssymb}
\usepackage{dsfont}
\usepackage{hyperref}
\usepackage{xcolor}

\usepackage{tikz}
\usetikzlibrary{arrows,positioning,calc}

\usepackage{namelabels}

\usepackage{enumerate}
\usepackage{booktabs}
\usepackage{rotating}

\usepackage[scaled=0.81]{beramono}  

\usepackage{listings}
\usepackage{xspace}

\usepackage[colorinlistoftodos,textwidth=10mm,textsize=tiny,obeyFinal]{todonotes}

\newcommand{\cut}{\ensuremath{\mathit{cut}}}

\definecolor{studentcol}{RGB}{27,158,119}
\definecolor{nondevelopcol}{RGB}{217,95,2}
\definecolor{ccol}{RGB}{117,112,179}

\newcommand{\student}[1]{{\textit{#1}}}
\newcommand{\devel}[1]{{\underline{#1}}}
\newcommand{\team}[1]{Team~\textsl{\nref{#1}}\xspace}
\newcommand{\pred}[1]{\textit{#1}}

\begin{document}

\title{VerifyThis 2019: A Program Verification Competition}
\iflong \subtitle{Extended Report} \fi

\titlerunning{VerifyThis 2019}

\author{Claire Dross \and
  Carlo A.\ Furia \\
  Marieke Huisman \and
  Rosemary Monahan \and
  Peter M\"uller
}

\authorrunning{Dross et al.}

\institute{Claire Dross \at
  AdaCore, France\\
  \email{dross@adacore.com}
  \and
  Carlo A. Furia \at
  USI Universit\`a della Svizzera italiana, Switzerland \\
  \email{furiac@usi.ch}
  \and
  Marieke Huisman \at
  University of Twente, the Netherlands \\
  \email{m.huisman@utwente.nl}
  \and
  Rosemary Monahan \at
  Maynooth University, Ireland\\
  \email{Rosemary.Monahan@mu.ie}
  \and
  Peter M\"uller \at
  ETH Zurich, Switzerland\\
  \email{Peter.Mueller@inf.ethz.ch}
}

\date{Received: date / Accepted: date}

\maketitle

\begin{abstract}
  VerifyThis is a series of program verification competitions that emphasize the human aspect:
  participants \linebreak tackle the verification of detailed behavioral properties---something that lies beyond the capabilities of fully automatic verification, and requires instead human expertise to suitably encode programs, specifications, and invariants.
  This paper describes the 8th edition of VerifyThis, which took place at ETAPS 2019 in Prague.
  Thirteen teams entered the competition, which consisted of three verification challenges and spanned two days of work.
  This report analyzes how the participating teams fared on these challenges,
  reflects on what makes a verification challenge more or less suitable for the typical VerifyThis participants,
  and outlines the difficulties of comparing the work of teams using wildly different verification approaches in a competition focused on the human aspect.
\end{abstract}


\section{The VerifyThis 2019 Verification Competition}
\label{sec:intro}

VerifyThis is a series of \emph{program verification competitions}
where participants prove expressive input/output properties of small
programs with complex behavior. This report describes VerifyThis 2019,
which took place on 6--7 April 2019
in Prague, Czech Republic, as a two-day event of the European Joint
Conferences on Theory and Practice of Software (ETAPS 2019).  It was
the eighth event in the series, after the VerifyThis competitions held
at FoVeOOS 2011, FM~2012, the Dagstuhl Seminar 14171 (in 2014), and
ETAPS 2015--2018.
The organizers of VerifyThis 2019
were also the authors of this paper---henceforth referred to as ``we''.

VerifyThis aims to bring together researchers and practitioners
interested in formal verification, providing them with an opportunity
for engaging, hands-on, and fun discussion. The results of the
competition help the research community evaluate progress and
assess the usability of formal verification tools in a controlled
environment---which still represents, on a smaller scale, important
practical aspects of the verification process.

Unlike other verification competitions that belong to the same TOOLympics
(Competitions in Formal Methods) track of ETAPS, VerifyThis emphasizes
verification problems that go beyond what can be proved fully
automatically, and require instead human experts ``in the loop''.
During a VerifyThis event, participating teams are given a number of
verification challenges that they have to solve on-site during the
time they have available using their favorite verification tools.  A
challenge is typically given as a natural-language description---possibly complemented with some pseudo-code or lightweight
formalization---of an algorithm and its specification.  Participants
have to implement the algorithm in the input language of their tool of
choice, formalize the specification, and formally prove the
correctness of the implementation against the specification.  The
challenge descriptions leave a lot of details open, so that
participants can come up with the formalization that best fits the
capabilities of their verification tool of choice. Correctness proofs
usually require participants to supply additional information, such as
invariants or interactive proof commands.

Following a format that consolidated over the years, VerifyThis 2019 proposed three verification challenges.
During the first day of the competition, participants worked during three 90-minute slots---one for each challenge.
Judging of the submitted solutions took place during the second day of the competition,
when
we assessed the level of correctness, completeness, and elegance of the submitted solutions.
Based on this assessment, we awarded prizes to the best teams in different categories (such as overall best team, and best student teams)
The awards were announced during the ETAPS lunch on Monday, 8 April 2019.

\iflong
\paragraph{Outline.}
The rest of this report describes VerifyThis 2019 in detail, and discusses the lessons we learned about the state of the art in verification technology.
\autoref{sec:challenges} outlines how we prepared the challenges; \iflong \autoref{sec:tutorial} discusses the invited tutorial that opened VerifyThis; \fi \autoref{sec:participants} presents the teams that took part in this year's VerifyThis;
and \autoref{sec:judging} describes the judging process in some more detail.

Then, Sections~\ref{sec:challenge1}--\ref{sec:challenge3} each describe a verification challenge in detail:
the content of the challenge, what aspects we weighed when designing it, how the teams fared on it, and a postmortem assessment of what aspects made the challenge easy or hard for teams.

Finally, \autoref{sec:discussion} presents the lessons learned from
organizing this and previous competitions, focusing on the tools and tool features that emerged,
on the characteristics of the challenges that made them more or less difficult for participants, 
and on suggestions for further improvements to the competition format.
\fi

The online archive of VerifyThis
\begin{center}
  \url{http://verifythis.ethz.ch}
\end{center}
includes the text of all verification challenges, and the solutions submitted by the teams (typically revised and improved after the competition).
Reports about previous editions of VerifyThis are also available~\cite{vt11,vt12,vt14,vt15,vt16,vt17,vt18}.
The motivation and initial experiences of organizing verification competitions in the style of VerifyThis are discussed elsewhere~\cite{vcomp-0,vcomp-organization};
a recent publication~\cite{vt-retrospective} draws lessons from the history of VerifyThis competitions.

\subsection{Challenges}
\label{sec:challenges}

A few months before the competition, we
sent out a public ``Call for Problems'' asking for suggestions of verification challenges that could be used during the competition.
Two people submitted by the recommended deadline proposals for three problems; and one more problem proposal arrived later, close to the competition date.

We combined these proposals with other ideas in order to design three challenges suitable for the competition.
Following our experience, and the suggestions of organizers of previous VerifyThis events, we looked for problems that were suitable for a 90-minute slot, and that were not too biased towards a certain kind of verification language or tool.
A good challenge problem should be presented as a series of specification and verification steps of increasing difficulty; even inexperienced participants should be able to approach the first steps, whereas the last steps are reserved for those with advanced experience in the problem's domain, or that find it particularly congenial to the tools they're using.
Typically, the first challenge involves an algorithm that operates on arrays or even simpler data types;
the second challenge targets more complex data structures in the heap (such as trees or linked lists);
and the third challenge involves concurrency.

In the end, we used one suggestion\ collected through the ``Call for Problems'' as the basis of the first challenge, which involves algorithms on arrays (see \autoref{sec:challenge1}).
Another problem suggestion\ was
the basis of the second challenge, which targets the construction of binary trees from a sequence of integers (see \autoref{sec:challenge2}).
For the third challenge, we took a variant of the matrix multiplication problem (which was already used, in a different form, during VerifyThis~2016)
that lends itself to a parallel implementation (see \autoref{sec:challenge3}).

\iflong
\subsection{Invited Tutorial}
\label{sec:tutorial}

We invited Virgile Prevosto to open VerifyThis 2019 with a tutorial about Frama-C\@.
Developed by teams at CEA LIST and INRIA Saclay in France, Frama-C\footnote{\url{https://frama-c.com}} is an extensible platform for source-code analysis of software written in C\@.

Frama-C works on C code annotated with specifications and other directives for verification written as comments in the ACSL (pronounced ``axel'') language.
Each plug-in in Frama-C provides a different kind of analysis, including classic dataflow analyses, slicing, and also dynamic analyses.
The tutorial\footnote{\url{https://frama.link/fc-tuto-2019-04}}
focused on the WP (Weakest Precondition) plugin, which supports deductive verification using SMT solvers or interactive provers to discharge verification conditions.

The tutorial began with the simple example of a function that swaps two pointers.
Despite the simplicity of the implementation, a complete correctness proof is not entirely trivial since it involves proving the absence of undefined behavior---a characteristic of C's memory model.
The tutorial continued with examples of increasing complexity demonstrating other features of the WP plugin and of the ACSL annotation language,
such as how to specify frame conditions and memory separation, how to reason about termination, and how to define and use custom predicates for specification.

Frama-C has been used to analyze critical low-level code,
such as the Contiki embedded operating system and implementations of critical communications protocols.
Its focus and the rich palette of analyses it supports make it a tool with an original approach to formal verification---one
that VerifyThis participants found interesting and stimulating to compare to the capabilities of their own tools.
\fi

\begin{table*}[tb]
  \centering
\newcounter{teamnum}
  \begin{tabular}{rlllr}
    & \multicolumn{1}{c}{\textsc{team name}} & \multicolumn{1}{c}{\textsc{members}} & \multicolumn{2}{c}{\textsc{tool}} \\
    \midrule
 \refstepcounter{teamnum}\theteamnum\namelabel{mergesort}{Merge\-sort} & \nref{mergesort}	& \student{Quentin Garchery} & Why3&\cite{why3,why3-2} \\
 \refstepcounter{teamnum}\theteamnum\namelabel{vercors}{Ver\-Cors T(w/o)o} & \nref{vercors} & \devel{Marieke Huisman}, \devel{Sebastiaan Joosten} & VerCors&\cite{vercors,vercors-2} \\
 \refstepcounter{teamnum}\theteamnum\namelabel{bashers}{Ba\-shers} & \nref{bashers} & \devel{Mohammad Abdulaziz}, \devel{\student{Maximilian P L Haslbeck}} & Isabelle&\cite{isabelle} \\
 \refstepcounter{teamnum}\theteamnum\namelabel{jm}{Jour\-dan-M\'e\-vel} & \nref{jm} & Jacques-Henri Jourdan, \student{Glen Mével} & Coq&\cite{coq-book,coq-idris} \\
 \refstepcounter{teamnum}\theteamnum\namelabel{openjml}{Open\-JML} & \nref{openjml} & \devel{David Cok} & OpenJML&\cite{openjml} \\
 \refstepcounter{teamnum}\theteamnum\namelabel{yvette}{Y\-VeT\-Te} & \nref{yvette} & \devel{Virgile Prevosto}, \devel{\student{Virgile Robles}} & Frama-C&\cite{framac} \\
 \refstepcounter{teamnum}\theteamnum\namelabel{refiners}{The Refiners} & \nref{refiners} & \devel{Peter Lammich}, \devel{\student{Simon Wimmer}} & Isabelle&\cite{isabelle,isabelle-2} \\
 \refstepcounter{teamnum}\theteamnum\namelabel{kiv}{KIV} & \nref{kiv} & \devel{\student{Stefan Bodenmüller}}, \devel{Gerhard Schellhorn} & KIV&\cite{kiv} \\
 \refstepcounter{teamnum}\theteamnum\namelabel{sw}{Sophie \& Wytse} & \nref{sw} & \devel{\student{Sophie Lathouwers}}, \devel{\student{Wytse Oortwijn}} & VerCors&\cite{vercors} \\
 \refstepcounter{teamnum}\theteamnum\namelabel{coinductive}{Coinductive Sorcery} & \nref{coinductive} & \student{Jasper Hugunin} & Coq&\cite{coq-book} \\
 \refstepcounter{teamnum}\theteamnum\namelabel{heja}{Heja mig} & \nref{heja} & \student{Christian Lidstr\"om} & Frama-C&\cite{framac} \\
 \refstepcounter{teamnum}\theteamnum\namelabel{eindhoven}{Eindhoven UoT} & \nref{eindhoven} & \devel{Jan Friso Groote}, \devel{\student{Thomas Neele}} & mCRL2&\cite{mcrl2,mcrl2-2} \\
 \refstepcounter{teamnum}\theteamnum\namelabel{viper}{Viper} & \nref{viper} & \devel{Alexander J. Summers} & Viper&\cite{viper}
  \end{tabular}
  \caption{Teams participating in VerifyThis 2019, listed in order of registration. For each \textsc{team} the table reports its \textsc{name}, its \textsc{members}, and the verification \textsc{tool} they used. A member names is in \student{italic} if the member is a student; and it is \devel{underlined} if the member is also a developer of the tool or of some extension used in the competition.}
  \label{tab:teams}
\end{table*}

\subsection{Participants}
\label{sec:participants}

\autoref{tab:teams} lists the thirteen teams that participated in VerifyThis 2019.
Four teams consisted of a single person, whereas the majority of teams included two persons (the maximum allowed).

As it is often the case during verification competitions, the majority of participants used a tool they know very well because they have contributed to its development. However, four teams identified themselves as non-developers, as
they did not directly contribute to the development
of the verification tools they used during the competition.

Out of 21 participants, 11 were graduate students. Some participated with
a senior colleague, while some others \linebreak worked alone or with other
students, making up a total of three all-student teams.

\subsection{Judging}
\label{sec:judging}

Judging took place on the competition's second day.
Each team sat for a 20--30-minute interview with us,
during which they went through
their solutions, pointing out what they did and didn't manage to verify,
and which aspects they found the most challenging.

Following the suggestions of previous organizers~\cite{vt-retrospective}, we asked teams to fill in a questionnaire about their submitted solutions
in preparation for the interview.
The questionnaire asked them to explain the most important features of the implementation, specification, and verification in their solutions,
such as whether the implementation diverged from the pseudo-code given in the challenge description,
whether the specification included properties such as memory safety,
and whether verification relied on any simplifying assumptions.
The questionnaire also asked participants to reflect on the process they followed (How much human effort was involved? How long would it take to complete your solution?), and on the strengths and weaknesses of the tools they used.
With the bulk of the information needed for judging available in the questionnaire, we could focus the interviews on the aspects that the participants found the most relevant while still having basic information about all teams.

At the same time as judging was going on,
participants not being interviewed were giving short presentations of their solutions to the other teams.
This is another time-honored tradition of VerifyThis, which contributes more value to the event and makes it an effective forum to exchange ideas about how to do verification in practice.
We briefly considered the option of merging interviews (with organizers) and presentation (to other participants),
but in the end we decided that having separate sessions makes judging more effective and lets participants discuss freely with others without the pressure of the competition---although the atmosphere was generally quite relaxed!

Once the interviews were over, we discussed privately to
choose the awardees. We structured our discussion around the
questionnaires' information, and sup\-ple-\linebreak mented it with the notes taken
during the interviews.  Nevertheless, we did not use any fixed
quantitative scoring, since VerifyThis's judging requires us to
compare very different approaches and solutions to the same
problems. Even criteria that are objectively defined in principle may
not be directly comparable between teams;
for example, correctness is relative to a specification, and
hence, different ways of formalizing a specification drastically change
the hardness of establishing correctness.
We tried to keep an open mind towards solutions that pursued an approach
very different from the one we had in mind when writing the challenges, provided
the final outcome was convincing.
Still, inevitably, our background, knowledge, and expectations somewhat may
have biased the judging process.  In the end, we were pleased by all
submissions, which showed a high level of effort, and results that
were often impressive---especially considering the limited available
time to prepare a solution.

We awarded six prizes in four categories:
\begin{itemize}
\item \emph{Best Overall Team} went to \team{refiners}
\item \emph{Best Student Teams} went to \team{mergesort} and \linebreak \team{sw}
\item \emph{Most Distinguished Tool Feature} went to \team{bashers}---for a library to model concurrency in Isabelle, which they developed specifically in preparation for the com\-pe\-ti\-tion---and to \team{vercors}---for their usage of ghost method parameters to model sparse matrices
\item \emph{Tool Used by Most Teams} went to Viper---used directly or indirectly\footnote{VerCors uses Viper as back-end; hence, \team{viper} used it directly, and \team{vercors} and \team{sw} used it indirectly.} by three different teams---rep\-re\-sented by Alexander J.\ Summers.
\end{itemize}

\section{Challenge 1: Monotonic Segments and GHC Sort}
\label{sec:challenge1}

The first challenge was based on the generic sorting algorithm
used in Haskell's GHC compiler.\footnote{\url{https://hackage.haskell.org/package/base-4.12.0.0/docs/src/Data.OldList.html\#sort}}
The algorithm is a form of \emph{patience sorting}.\footnote{Named after the patience card game \url{https://en.wikipedia.org/wiki/Patience_sorting}.}

\subsection{Challenge Description}
\label{ch1:description}

\iflong
Challenge~1 was in two parts---described in \autoref{sec:challenge1A} and \autoref{sec:challenge1B}---each consisting of several different verification tasks.
We did not expect participants to solve both parts in the 90 minutes at their disposal, but suggested that they
pick the one that they found the most feasible given the tool they were using and their preferences.
\fi

\subsubsection{Part A: Monotonic Segments}
\label{sec:challenge1A}

Given a sequence $s$
\[
  s\quad=\quad s[0]\,s[1]\,\ldots\,s[{n-1}]\qquad n \geq 0
\]
of elements over a totally sorted domain (for example, the integers),
we call \textbf{monotonic cutpoints} any indexes that cut $s$ into segments that are monotonic: 
each segment's elements are all increasing or all decreasing.\footnote{More precisely, all strictly increasing, or nonincreasing (decreasing or equal).}
Here are some examples of sequences with monotonic cutpoints:
\begin{center}
  \scriptsize
  \begin{tabular}{rrc}
    \textsc{sequence} $s$ & \textsc{monotonic cutpoints} & \textsc{monotonic segments} \\
    \midrule
    $1\ 2\ 3\ 4\ 5\ 7$ & $0\ 6$ & $1\ 2\ 3\ 4\ 5\ 7$ \\
    $1\ 4\ 7\ 3\ 3\ 5\ 9$ & $0\ 3\ 5\ 7$ & $1\ 4\ 7 \mid 3\ 3 \mid 5\ 9$\\
    $6\ 3\ 4\ 2\ 5\ 3\ 7$ & $0\ 2\ 4\ 6\ 7$ & $6\ 3 \mid 4\ 2 \mid 5\ 3\mid 7$
  \end{tabular}
\end{center}
In this challenge we focus on \textbf{maximal} monotonic cutpoints, that is such that, if we extend any segment by one element, the extended segment is not monotonic anymore.

\iflong
Formally, given a sequence $s$ as above, we call \textbf{monotonic cutpoints} \emph{any} integer sequence
\[
  \cut\quad=\quad c_0\,c_1\,\ldots\,c_{m-1}
\]
such that the following four
properties hold:\begin{description}[within bounds:]
\item[\emph{non-empty:}] $m > 0$
\item[\emph{begin-to-end}:] $c_0 = 0$ and $c_{m-1} = n$
\item[\emph{within bounds}:] for every element $c_k \in \cut$: $0 \leq c_k \leq n$
\item[\emph{monotonic}:] for every pair of consecutive elements \linebreak $c_k, c_{k+1} \in \cut$, the segment
    $s[c_k..c_{k+1}) = s[{c_k}]\,s[{c_k +1}]\,\ldots\,s[{c_{k+1}-1}]$
    of $s$, which starts at index $c_k$ included and ends at index $c_{k+1}$ excluded, is \emph{monotonic}, that is:
    either $s[{c_k}] < s[{c_k +1}] < \cdots < s[{c_{k+1}-1}]$ or $s[{c_k}] \geq s[{c_k +1}] \geq \cdots \geq s[{c_{k+1}-1}]$
    \label{eq:monotonic-def}
\end{description}
\fi

Given a sequence $s$, for example stored in an array, maximal monotonic cutpoints can be computed by scanning $s$ once while storing every index that corresponds to a change in monotonicity (from increasing to decreasing, or vice versa), as shown by the algorithm in \autoref{lst:cutpoints-algo}.
\begin{figure}[tb]
  \centering
\begin{lstlisting}[xleftmargin=0mm]
  cut := [0]  # singleton sequence with element 0
  x, y := 0, 1
  while y < n:   # n is the length of sequence s
     increasing := s[x] < s[y]  # in increasing segment?
     while y < n and (s[y-1] < s[y]) == increasing:
        y := y + 1
     cut.extend(y)  # extend cut by adding y to its end
     x := y
     y := x + 1
  if x < n:
     cut.extend(n)
\end{lstlisting}
  \caption{Algorithm to compute the maximal cutpoints \lstinline|cut| of sequence~\lstinline|s|.}
  \label{lst:cutpoints-algo}
\end{figure}

To solve Challenge~1.A, we asked participants to carry out the following tasks.

\begin{description}
\item[\textbf{Implementation}] task:
  Implement the algorithm in \autoref{lst:cutpoints-algo} to compute monotonic cutpoints of an input sequence.
  
  \item[\textbf{Verification}] tasks:
    \begin{enumerate}
    \item Verify that the output sequence \iflong satisfies properties \emph{non-empty}, \emph{begin-to-end}, and \emph{within bounds} above.\fi

    \item Verify that \iflong the output sequence  satisfies property \emph{monotonic} given above (\emph{without} the maximality requirement).\fi

    \item Strengthen the definition of monotonic cutpoints so that it requires \emph{maximal} monotonic cutpoints, and prove that your algorithm implementation computes maximal cutpoints according to the strengthened definition.
    \end{enumerate}
\end{description}

\subsubsection{Part B: GHC Sort}
\label{sec:challenge1B}

To sort a sequence~$s$, \textbf{GHC Sort} works as follows:

\begin{enumerate}
\item Split $s$ into monotonic segments $\sigma_1, \sigma_2,\ldots,\sigma_{m-1}$
\item Reverse every segment that is decreasing
\item Merge the segments \emph{pairwise} in a way that preserves the order
\item If all segments have been merged into one, that is an ordered copy of $s$; then terminate.
  Otherwise, go to step~3
\end{enumerate}
Merging in step~3 works like merging in Merge Sort\iflong, which follows the algorithm in \autoref{lst:merge-algo}\fi.\iflong
\begin{figure}[tb]
  \centering
\begin{lstlisting}[xleftmargin=7mm]
# merge ordered segments s and t
merged := []
x, y := 0, 0
while x < length(s) and y < length(t):
  if s[x] < t[y]:
    merged.extend(s[x])
    x := x + 1
  else:
    merged.extend(t[y])
    y := y + 1
# append any remaining tail of s or t
while x < length(s):
  merged.extend(s[x])
  x := x + 1
while y < length(t):
  merged.extend(t[y])
  y := y + 1
\end{lstlisting}
  \caption{Algorithm to merge sorted sequences \lstinline|s| and \lstinline|t| into sorted sequence \lstinline|merged|.}
  \label{lst:merge-algo}
\end{figure}

For example, GHC Sort applied to the sequence $s = 3\ 2\ 8\ 9\ 3\ 4\ 5$ goes through the following steps:
\begin{itemize}
\item monotonic segments: $3\ 2\mid 8\ 9\mid 3\ 4\ 5$
\item reverse decreasing segments: $2\ 3 \mid 8\ 9\mid 3\ 4\ 5$
\item merge segments pairwise: $2\ 3\ 8\ 9 \mid 3\ 4\ 5$
\item merge segments pairwise again: $2\ 3\ 3\ 4\ 5\ 8\ 9$, which is $s$ sorted
\end{itemize}
\fi

To solve Challenge~1.B, we asked participants to carry out the following tasks.

\begin{description}
 \item[\textbf{Implementation}] task:
  Implement GHC Sort in your programming language of choice.
  
  \item[\textbf{Verification}] tasks:
    \begin{enumerate}
    \item Write functional specifications of all procedures/functions/main steps of your implementation.
    \item Verify that the implementation of \emph{merge} returns a sequence \lstinline|merged| that is \textbf{sorted}.
    \item Verify that the overall sorting algorithm returns an output that is sorted.
    \item Verify that the overall sorting algorithm returns an output that is a permutation of the input.
    \end{enumerate}
\end{description}

\iflong
\subsection{Designing the Challenge}
\label{ch1:design}

The starting point for designing this challenge
was Nadia Polikarpova's suggestion to target GHC's generic sorting
method.  Responding to VerifyThis's Call for Problems, she submitted a
concise high-level description of how the sorting algorithm works, and
pointed us to an implementation in Liquid
Hask\-ell\footnote{\url{https://github.com/ucsd-progsys/liquidhaskell/blob/develop/tests/pos/GhcSort1.hs}}
that verifies sortedness of the output.

In order to understand whether this algorithm could be turned into a
suitable verification challenge, we developed a prototype
implementation of GHC Sort written in Python, complete with assertions
of key correctness properties as well as tests that exercised the
implementation on different inputs.  Tweaking this implementation was
useful to quickly explore different variants of the algorithm and their
repercussions on correct program behavior.

We also developed a verified Dafny implementation of parts of the
algorithm, in order to get an idea of the kinds of invariants that are
required for proving correctness and to anticipate possible pitfalls
when trying to specify or verify the algorithm.

These attempts indicated that verifying the whole GHC Sort algorithm
would have been a task too demanding for a 90-minute slot. Therefore,
we split it into two conceptually separate parts: A) finding the
monotonic segments of the input (\autoref{sec:challenge1A}); and B) the
actual sorting procedure (\autoref{sec:challenge1B}). We suggested to
participants to focus their work on the parts of the algorithm that
were more amenable to analysis according to the capabilities of their
verification tool, while specifying the expected behavior of the other
parts without proving their correctness explicitly. In particular, to
decouple the different parts of the challenge and give more
flexibility, we left participants working on part~B free to add the
reversal (step~2 of GHC Sort) to the same pass that constructs the
monotonic segments in step~1.

GHC Sort's original implementation is in Haskell---a pure
functional programming language, which offers abstract lists as a
native data type---bringing the risk of a verification challenge
biased in favor of tools based on functional programming features. To
mitigate this risk, we explicitly told participants they were free to
choose any representation of input sequences and cutpoints sequences
that was manageable using their programming language of choice:
arrays, mathematical sequences, dynamic lists, \ldots. We also
presented the key algorithms (\autoref{lst:cutpoints-algo} and
\autoref{lst:merge-algo}) using iteration, but still left participants free
to use recursion instead of looping to implement the general idea
behind the algorithms. 

One technical issue we discussed while preparing the challenge was the definition of
\emph{monotonicity} of a segment.  Definition~\emph{monotonic} on page~\pageref{eq:monotonic-def}
above is asymmetric since it distinguishes between strictly increasing
and nonstrictly decreasing (that is, nonincreasing) segments. While
using a symmetric definition---which would allow repeated equal
values to appear indifferently in increasing or decreasing segments---seemed more elegant and perhaps more natural, the asymmetric
definition \eqref{eq:monotonic-def} seemed simpler to implement,
since it is enough to compare the first two elements of a segment to
know whether the rest of the segment has to be increasing (strictly)
or decreasing (nonstrictly). In turn, definition
\eqref{eq:monotonic-def} seemed to require slightly simpler invariants
because the predicate for ``decreasing'' would be exactly the
complement of the predicate for ``increasing''.  At the same time, we
were wary of how people used to different notations and verification
styles might still find the symmetric definition easier to work with.
Therefore, we left participants free to change the definition of
\emph{monotonic} so that segments of equal values could be
indifferently included in increasing or in decreasing segments. If
they choose to do so, we also pointed out that they may have had to
change the algorithm in \autoref{lst:cutpoints-algo} to match their
definition of monotonic segment.

One final aspect that we tried to anticipate was the requirement of
\emph{maximality} of the monotonic segments.  Proving maximality
seemed somewhat more complex than proving monotonicity alone; hence,
we marked it as ``optional task (advanced)'' and we did not provide
any formal definition of maximality---so that participants were free
to come up with the formal specification that best fitted their
general solution.
\fi

\subsection{Submitted Solutions}
\label{ch1:results}

\iflong
\subsubsection*{Overall Results}
\fi
\team{openjml} and \team{refiners} submitted solutions of challenge~1 that were complete
and correct.  Another team got close but missed a few crucial
invariants.  Five teams made substantial progress but introduced some
simplifying assumptions or skipped verification of maximality.  And
another five teams' progress was more limited, often due to a mismatch
between their tools' capabilities and what was required by the
challenge.

\iflong
\subsubsection*{Detailed Results}
The two teams using Isabelle followed very different approaches to
representing cutpoints in challenge~1.  While \linebreak \team{refiners} used
functional lists of lists to represent monotonic segments explicitly,
\team{bashers} chose to use an explicit representation of indexes
corresponding to cutpoints---which turned out not to be a good match
for Isabelle's functional programming features.  \team{refiners}
expressed challenge~1's correctness properties recursively to be
amenable to inductive proofs. With these adjustments, they could take
full advantage of Isabelle's verification capabilities: they specified
all properties of part~A and performed all verification tasks with the
exception of completing the proof of maximality; and they even managed
to solve most of part~B's specification and verification tasks,
completing all its proofs not long after the competition slot was over.

Both teams using the Coq theorem prover encoded challenge~1-A in a
purely functional setting, using lists and
recursion.  Without the support of domain-specific libraries, reasoning about the
properties required by the challenge turn\-ed out to be quite cumbersome
and time-consuming. In particular, Coq's constructive logic requires
that every recursive function definition be accompanied by a proof of
termination (showing that recursion is well founded). This slowed
down the work of \team{jm} and \team{coinductive}, who could 
submit only partial solutions in time for the competition.

Challenge~1---in particular, part A---was well-suited, in its
original form using arrays, with OpenJML's capabilities:
\team{openjml} delivered an implementation of the algorithms that was
very close to the pseudo-code of \autoref{lst:cutpoints-algo}, and
could express and prove properties that directly translated all of the
challenge's verification tasks.  As usual for verifiers based on SMT
solvers, a successful proofs depends on being able to write
specifications in a form amenable to automated
reasoning. Then, the required loop invariants had a fairly clear
connection to the postconditions that had to be proved. To save time,
\team{openjml} took some shortcuts in the implementation (for example,
writing the result into a global variable instead of returning it
explicitly) that do not affect its behavior but are somewhat
inelegant; cleaning them up, however, should be straightforward.

Both teams using VerCors progressed quite far in solving part~A of
challenge~1, but could not complete the proof of maximality during the
competition. \iflong\else\linebreak\fi \team{sw} modified the implementation of the algorithm to
compute the cutpoints so that it stores in a separate array the
monotonicity direction of each segment (that is whether each segment
is increasing or decreasing); this helped to simplify reasoning about
maximality, since one can more easily refer to the monotonicity of
each segment independent of the others. Even without this trick,
\team{vercors} progressed further in the proof of maximality, as they
only missed a few key invariants.
Both teams using VerCors used immutable sequences, instead of arrays,
to store cutpoint sequences; this dispensed them with having to deal
with per\-mis\-sions---extensively used for arrays by VerCors.

\team{kiv} also used immutable sequences as primary data structure for
challenge~1-A; KIV's libraries recently included a proof that
sequences and arrays can simulate each other, and hence, it should be
possible to rework the formalization to work with arrays with limited
changes. As it is customary in KIV, and in contrast to what most other
approaches prefer to do, \team{kiv} expressed all correctness
properties together using a single descriptive predicate. According to
\team{kiv}'s members, this helps scalability with their tool, but may
hamper a partial yet faster progress when limited time is available---as it was the case during the competition, when they could not
complete the proofs in time.

\team{viper} implemented challenge~1-A's algorithm using arrays; more
precisely, they introduced a \emph{domain definition} that represents
arrays as objects with certain properties.  \team{viper} modified the
algorithm in \autoref{lst:cutpoints-algo} trying to enforce the
property that increasing and decreasing segments strictly alternate---a property that the original algorithm does not possess.  This turned
out to be tricky to do and complicated several aspects of the
specification.  In the end, \team{viper} submitted a solution that
included several parts of the specification and invariants necessary
to prove correctness but did not completely establish monotonicity and
maximality.

\team{yvette} solved challenge~1-A using Frama-C's WP plugin, which
provides automated deductive verification of C code using SMT
solvers. Since Frama-C encodes low-level aspects of the C memory
model, correctness proofs often generate a large number of proof
obligations that require to establish safety and separation of
different memory regions. These low-level proof obligations may
significantly complicate the proof of higher-level functional
properties---such as those that are the main focus of VerifyThis's
challenges. More practically, this interplay of user-defined
predicates and low-level properties made Frama-C's WP plugin generate
proof obligations that were not automatically provable by SMT solvers
and would have required a lengthy manual analysis using an interactive
prover like Coq.  Due to these hurdles, \team{yvette} managed to get
close to a proof of monotonicity, but could not complete some
invariants and lemmas in time during the competition.

The only team using a model checker, \team{eindhoven} had to introduce
restrictions and simplification to express the requirements of
challenge~1-A within the finite-state expressiveness of their
verification tool.  In their solution, the integers that make up a
sequence range over a finite bound; and only input lists of a certain
fixed length could be analyzed.  In practice, most of their analysis
used lists of up to 4 elements (lists of up to 10 elements is close to
the maximum the tool can handle before the analysis algorithm exhausts the available resources); and they did not prove maximality (possibly because expressing
the property in operational form would have been tedious).

\subsection{Postmortem Evaluation of the Challenge}
\label{ch1:post-eval}
\fi

Teams did not find the definition \iflong\eqref{eq:monotonic-def}\fi of
monotonicity hard to work with because it is asymmetric: as far as we
could see, most of them encoded the property as we suggested and made
it work effectively.

However, a couple of teams were confused by mistakenly assuming a
property of monotonic segments: since the condition for ``decreasing''
is the complement of the condition for ``increasing'', they concluded
that increasing and decreasing segments must strictly alternate (after
a decreasing segment comes an increasing one, and vice versa).  This
is not true in general, as shown by the example of sequence
$6\ 3\ 4\ 2\ 5\ 3\ 7$, which is made of 4 monotonic segments
$6\ 3 \mid 4\ 2 \mid 5\ 3\mid 7$, all of them decreasing.

While we did not give a formal definition of maximality, the teams
that managed to deal with this advanced property did not have trouble
formalizing it. Since ``extending'' a segment can be generally done both on its right and on its left endpoint, teams typically expressed maximality as two separate properties: to the right and to the left. While it may be possible to prove that one follows from the other (and the definition of monotonic cutpoints), explicitly dealing with both variants was found to be preferable in practice since the invariants to prove one variant are clearly similar to those to prove the other.

\section{Challenge 2: Cartesian Trees}
\label{sec:challenge2}

The second challenge involved the notion of Cartesian trees\footnote{\url{https://en.wikipedia.org/wiki/Cartesian_tree}} of a sequence of integers and, in particular, dwelt on how such trees can be constructed in linear time from the sequence of all nearest smaller values\footnote{\url{hhttps://en.wikipedia.org/wiki/All_nearest_smaller_values}} of the input sequence.

\subsection{Challenge Description}
\label{ch2:description}

\iflong
This challenge was in two parts. The first part, presented in \autoref{sec:challenge2A}, asked to compute the sequence of all nearest smaller values of an input sequence, while the second, in \autoref{sec:challenge2B}, dealt with the construction of the sequence's actual Cartesian tree. \iflong
We did not expect participants to complete the whole challenge in an hour and a half; so they could choose the part that best fitted their language of choice. The second part of the challenge used features described in the first part, but participants did not need to actually implement and verify the algorithms of the first part to carry out the second. \fi
\fi
\subsubsection{Part A: All Nearest Smaller Values}
\label{sec:challenge2A}

For each index in a sequence of values, we define the nearest smaller value to the left, or left neighbor, as
the last index among the previous indexes that contains a smaller value. \iflong More precisely,
for each index $x$ in an input sequence $s$, the \emph{left neighbor} of $x$ in $s$ is the index $y$ such that:

\begin{itemize}
 \item $y<x$,
 \item the value stored at index $y$ in $s$, written $s[y]$, is smaller than the value stored at index $x$ in $s$,
 \item there are no other values smaller than $s[x]$ between $y$ and~$x$.
\end{itemize}
\fi
There are indexes that do not have a left neighbor; for example, the first value, or the smallest value in a sequence.

We consider here an algorithm that constructs the sequence of left neighbors of all values of a sequence $s$. \iflong
It works using a stack. At the beginning, the stack is empty. Then, for each index $x$ in the sequence, pop indexes from the stack until a index $y$ is found such that $s[y]$ is smaller than $s[x]$. If such a index exists in the stack, it is the left neighbor of $x$; otherwise, $x$ does not have a left neighbor. After processing $x$, push $x$
onto the stack and go to the next index in $s$. \fi This algorithm is given in pseudo-code in \autoref{lst:smaller-nearest-algo}.

\begin{figure}[tb]
\begin{lstlisting}
stack := []  # empty stack
for every index x in s:
    # pop values greater or equal to s[x]
    while not stack.is_empty
          and s[stack.top] >= s[x]:
        stack.pop
            
    if stack.is_empty:
        # x doesn't have a left neighbor
        left[x] := 0
    else:
        left[x] := stack.top

    stack.push (x)
\end{lstlisting}
  \caption{Algorithm to compute the sequence \lstinline|left| of all left nearest smaller values of input sequence \lstinline|s|. The algorithm assumes that \textbf{indexes start from $1$}, and hence, it uses $0$ to denote that an index has no left neighbor. }
  \label{lst:smaller-nearest-algo}
\end{figure}

As an example, consider sequence $s =4\, 7\, 8\, 1\, 2\, 3\, 9\, 5\, 6$. The sequence of the left neighbors of $s$ (using indexes that start from 1) is: $\texttt{left} = 0\, 1\, 2\, 0\, 4\, 5\, 6\, 6\, 8$. The left neighbor of the first value of $s$ is $0$ (denoting no valid index), since the first value in a list has no values at its left. The fourth value of $s$ (value $1$) is also $0$, since $1$ is the smallest value of the list.

To solve Challenge~2.A, we asked participants to carry out the following tasks:

\begin{description}
\item[\textbf{Implementation}] task.
Implement the algorithm to compute the sequence of left neighbors from an input sequence.

\item[\textbf{Verification}] tasks.
\begin{enumerate}
 \item \emph{Index}: verify that, for each index $i$ in the input sequence $s$, the left neighbor of $i$ in $s$ is smaller than $i$, that is $\texttt{left} [i] < i$.
 \item \emph{Value}: verify that, for each index $i$ in the input sequence $s$, if $i$ has a left neighbor in $s$, then the value stored in $s$ at the index of the left neighbor is \linebreak smaller than the value stored at index $i$, namely, if $\texttt{left} [i]$ is a valid index of $s$ then $s [\texttt{left} [i]] < s [i]$.
 \item \emph{Smallest}: verify that, for each index $i$ in the input sequence $s$, there are no values smaller than $s[i]$ between $\texttt{left} [i] + 1$ and $i$ (included).
\end{enumerate}
\end{description}

\subsubsection{Part B: Construction of a Cartesian Tree}
\label{sec:challenge2B}

\begin{figure}[!tb]
  \centering
  \begin{tikzpicture}[
    treenode/.style={draw,circle,very thick,
                     minimum width=6mm,inner sep=1mm,outer sep=0mm}
    ]
    \matrix[row sep=0pt, column sep=2mm] {
      \node(s4){$4$}; & \node(s7){$7$}; & \node(s8){$8$}; & \node(s1){$1$}; & \node(s2){$2$}; & \node(s3){$3$}; & \node(s9){$9$}; & \node(s5){$5$}; & \node(s6){$6$}; \\[5mm]
      &&& \node[treenode] (n1) {$1$}; \\
      &&&& \node[treenode] (n2) {$2$}; \\
      &&&&& \node[treenode] (n3) {$3$}; \\
      \node[treenode] (n4) {$4$}; \\
      &&&&&&& \node[treenode] (n5) {$5$}; \\
      &&&&&&&& \node[treenode] (n6) {$6$}; \\
      & \node[treenode] (n7) {$7$}; \\
      && \node[treenode] (n8) {$8$}; \\
      &&&&&& \node[treenode] (n9) {$9$}; \\
    };
    \begin{scope}[-latex,very thick]
      \draw (n1) -- (n2);
      \draw (n2) -- (n3);
      \draw (n1) -- (n4);
      \draw (n3) -- (n5);
      \draw (n4) -- (n7);
      \draw (n7) -- (n8);
      \draw (n5) -- (n6);
      \draw (n5) -- (n9);
    \end{scope}
    \foreach \n in {1, 2, ..., 9} \draw[thin,dotted] (s\n) -- (n\n);
  \end{tikzpicture}
\caption{Cartesian tree of sequence $4\ 7\ 8\ 1\ 2\ 3\ 9\ 5\ 6$.}
  \label{fig:cartesian-tree}
\end{figure}
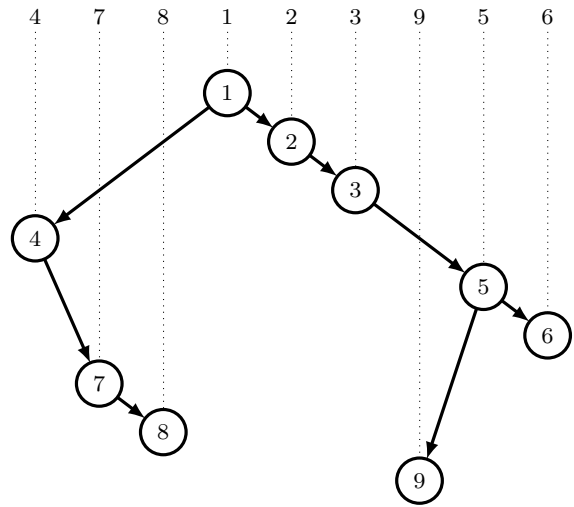

Given a sequence $s$ of \emph{distinct} numbers, its unique \emph{Cartesian tree} $CT(s)$ is the tree such that:
\begin{enumerate}
 \item $CT(s)$ contains exactly one node per value of $s$.
 \item When traversing $CT(s)$ in-order---that is, using a symmetric traversal: first visit the left subtree, then the node itself, and finally the right subtree---el\-e\-ments are encountered in the same order as $s$.
 \item Tree $CT(s)$ has the heap property---that is, each node in the tree contains a value (not an index) bigger than its parent's.
\end{enumerate}
The Carthesian tree of sequence $s =4\, 7\, 8\, 1\, 2\, 3\, 9\, 5\, 6$ is given in \autoref{fig:cartesian-tree}.

\iflong
There are several algorithms to construct a Cartesian tree in linear time from its input sequence. The one we consider here is based on the all nearest smaller values problem (part A of this challenge). Let's consider a sequence of distinct numbers $s$. First, we \fi construct the sequence of left neighbors for the value of $s$ using the algorithm in \autoref{lst:smaller-nearest-algo}. Then, we construct the sequence of right neighbors using the same algorithm, but starting from the end of the list. Thus, for every index $x$ in sequence $s$,
the parent of $x$ in $CT(s)$ is either:
\begin{itemize}
 \item The left neighbor of $x$ if $x$ has no right neighbor.
 \item The right neighbor of $x$ if $x$ has no left neighbor.
 \item If $x$ has both a left neighbor and a right neighbor, then $x$'s parent is the larger one.
 \item If $x$ has no neighbors, then $x$ is the root node.
\end{itemize}

To solve Challenge~2.B, we asked participants to carry out the following tasks:

\begin{description}
 \item[\textbf{Implementation}] task.
Implement the algorithm for the construction of the Cartesian tree.

\item[\textbf{Verification}] tasks.
\begin{enumerate}
 \item \emph{Binary:} verify that the algorithm returns a well \linebreak formed binary tree, with one node per value (or per index) in the input sequence.
 \item \emph{Heap:} verify that the resulting tree has the heap property, that is, each non-root node contains a value \linebreak larger than its parent.
 \item \emph{Traversal:} verify that an in-order traversal of the tree traverses values in the same order as in the input sequence.
\end{enumerate}
\end{description}

\iflong
\subsection{Designing the Challenge}
\label{ch2:design}
The subject for the challenge was given to us by Gidon Ernst (one of the organizers of VerifyThis 2018) as an idea that was considered but, in the end, not used for the 2018 verification competition.

After first reading about Cartesian trees, we were wary of the risk that using them as subject would lead to a challenge too much oriented toward functional programming---unfeasible using verification tools that cannot handle recursive data structures such as trees and lists. To avoid this risk, we focused the challenge on one specific imperative algorithm that constructs a Cartesian tree bottom-up, attaching the nodes to their parents in the order in which they appear in the input sequence.

To better understand if we could make a challenge out of the this bottom-up Cartesian tree construction algorithm, we tried to implement and verify it using the SPARK verification tool for Ada. We began by writing and annotating the short loops that build the input sequence's nearest smaller values to the left and to the right. This task was not complicated, but turned out to be time-consuming enough to serve as a challenge by itself. Completing the implementation and verification of the actual Cartesian tree construction algorithm turned out to be decidedly more complicated: writing the algorithm itself was no big deal, but understanding how it works well enough to prove it correct was more challenging. In particular, proving property \emph{traversal} (in-order traversal of a Cartesian tree gives the input sequence) took nearly one day of work for a complete working solution in SPARK. 

Following these investigations, we considered the possibility of simply dropping from the challenge the construction of Cartesian trees, and concentrating only on the construction of nearest smaller values. However, we decided against that option, because we still wanted to give participants who had the right background and tools a chance of trying their hands at proving this challenging algorithm. To make the overall challenge tractable, we split it in two parts.

The first part, concerned only with nearest smaller values, was explicitly presented as the simplest, and was designed to be verifiable using a wide range of tools, at it only deals with sequences. Since the main algorithm (\autoref{lst:smaller-nearest-algo}) is imperative but uses stacks---which could make it a bit tricky to verify using only functional data structures---we let participants free to use an existing implementation of stacks or even use sequences as models of stacks.

As for the second part, dealing with the Cartesian tree construction algorithm, we clearly split the verification job in three distinct tasks of different difficulties; and marked the third task (property \emph{traversal}) as ``optional'', assuming that it would be mostly useful as a further exercise to be done after the competition.
We did not provide an algorithm in pseudo-code for this part, as writing an implementation is straightforward from the textual description but also depends strongly on the data structures used to encode the tree. Instead, we presented an example of a Cartesian tree built from a sequence, so that participants could use it to test their implementation and to understand why it worked.
We also remarked to the participants that they could implement trees as they preferred, using for example a recursive data-type, a pointer-based structure, or even just a bounded structure inside an array.

\fi

\subsection{Submitted Solutions}
\label{ch2:results}

\iflong
\subsubsection*{Overall Results} \fi
Two teams submitted solutions to challenge~2 that were both correct and complete: \team{openjml} worked on part~A of the challenge, and \team{vercors} on part~B. The latter team even managed to verify a partial specification of part~B's task \emph{traversal}---which was marked ``optional''. Another four teams completed the first two verification tasks of part~A, one of them coming close to finishing the proof of the third, with only a small part of the necessary invariant missing. Another team completed all three verification tasks of part A but with simplifying assumptions (on the finite range of inputs). Another two teams completed part~A's verification task~1 only.
The remaining four teams didn't go further than implementing the algorithm of the same part and writing partial specifications of the properties that were to be verified.

\iflong
\subsubsection*{Detailed Results}

Most teams attempted part~A of challenge~2, as it was presented as the more approachable of the two. Only two teams attempted part~B: \team{vercors}, using VerCors, who focused entirely on part B, and \team{refiners}, using Isabelle, whose two members decided to work separately in parallel---one person on each part of the challenge---to assess which was more feasible (and eventually decided to focus on part~A).

Both teams working on part~B represented trees using a ``parent'' relation mapping an index in the input sequence to the index of its parent node in the tree. 
\team{refiners} encoded this relation as a function on indexes. They managed to verify the second verification task (\emph{heap}: the tree is a heap), but then decided to continue to work on part~A of the challenge, since it seemed more suitable for their tool's capabilities.
In contrast, \team{vercors} stored the parent of each value in the input sequence using another sequence. They also defined two other arrays, storing the left and right child of each node. On tree structures encoded using this combination of parent and child relations,
\team{vercors} managed to complete part~B's verification tasks~1 and~2. They even verified a partial version of task~3's property \emph{traversal}---partial because it involved only a node's immediate children instead of the whole left and right subtrees.

Even though they tackled the same problem, the two submissions in Isabelle for part~A of the challenge were very different. \team{bashers} sticked to the usual functional programming style most common in Isabelle. They implemented the algorithm using two recursive functions to represent the two loops in the pseudo-code of \autoref{lst:smaller-nearest-algo}. By contrast, \team{refiners}---true to their name---deployed Isabelle's refinement framework to encode the algorithm directly in an iterative fashion, so that their implementation could closely match the pseudo-code in \autoref{lst:smaller-nearest-algo}. On top of this, they attempted refinement proofs to express part~A's three verification tasks. This worked well for the first two tasks (\emph{index} and \emph{value}), but they could not carry out the third one (\emph{smallest}) in time. While revising their solution after the competition, they realized that they had not implemented the algorithm correctly, because their encoding implied that no values in the input sequence can have a smaller value to its left. In principle, this mistake in the implementation should not have invalidated their proofs of verification tasks~1 and~2, which were expressed as conditionals on any values that do have smaller values to their left. Thus, once they noticed the error, they fixed the implementation and tried replaying the mechanized proofs of the first two properties. Even though they were using Sledgehammer to automate part of the reasoning, only the first task could be verified without manually adjusting the interactive proofs---which required some different proofs steps even though the overall proof logic was unchanged.

Both teams using Coq, \team{jm} and\linebreak \team{coinductive}, implemented a functional version of the pseudo-code in \autoref{lst:smaller-nearest-algo} using two recursive functions instead of loops---just like \team{bashers} did in Isabelle. This encoding proved tricky to get right: both teams ended up with a slightly incorrect ``off-by-one'' version of the algorithm that also pops (instead of just inspecting it) the first value \lstinline|y| on the stack that satisfies \lstinline|s[y] < s[x]| (exit condition of the inner loop in \autoref{lst:smaller-nearest-algo})
and thus is the left neighbor of current value \lstinline|x|.
This mistake does not affect the verification of tasks~1 and~2 (\emph{index} and \emph{value}),
and, in fact, the Coq teams did not notice it and still managed to specify (both teams) and prove (\team{jm}) these two tasks.
In contrast, the invariant needed to prove the third verification task (\emph{smallest}) depends on all values previously processed during the computation, which means that it could not have been expressed on the implementations written by the Coq teams but would have required additional information about processed values to be passed as part of the recursive functions' arguments.

As presented in \autoref{lst:smaller-nearest-algo}, the algorithm for the construction of the sequence of all nearest smaller values of an integer sequence was more suited to an imperative implementation. The Java implementation produced by \team{openjml} was indeed very close to that pseudo-code algorithm. It included a low-level stack implementation consisting of an array along with a separate variable storing the stack's top value index. The three properties---corresponding to the three verification tasks \emph{index}, \emph{value}, and \emph{smallest}---were expressed in a direct way, and all were verified automatically by OpenJML without manual input other than the suitable loop invariants. The loop invariant for the third verification task was by far the most complex, but, once it was expressed correctly, the automated prover~Z3---used as the backend of OpenJML---could handle it without difficulties in the automated proofs.

Other teams using a language with support for imperative programming features were also able to go quite far in the implementation and the verification of the algorithm of challenge~2's part~A. These submitted solutions' implementations closely matched the algorithm in \autoref{lst:smaller-nearest-algo} with differences only in how stacks were represented. \team{mergesort}, using Why3, encoded stacks as lists with an interface to query the first value (top) and retrieve the tail of the list (pop). The main limitation of this approach was the background solver's limited support for recursive lists. As a result, some of the lemmas about stacks required to build the algorithm's overall correctness proofs couldn't be verified automatically, and were left unproved in the submitted solution. Despite this issue, \team{mergesort} managed to verify the first two verification tasks, and made significant progress on the third one. The invariants submitted for this task were proved automatically and close to the required ones---even though they were not strong enough to complete the verification of task~\emph{smallest}.

\team{viper} also came close to a complete solution of part~A. The team's implementation of the algorithm was close to \autoref{lst:smaller-nearest-algo}'s, whereas the representation of stacks was more original. Instead of using a concrete data structure, \team{viper} defined stacks in a pure logic fashion using uninterpreted function symbols and axioms that postulate the result of popping, pushing, and peeking on a stack. \linebreak
\team{viper}'s submitted solution included specifications of all three verification tasks, and complete proofs of the first two.
Since the axiomatic representation did not support referencing arbitrary values inside the stack, \team{viper} resorted to expressing the invariant for the third verification task using a recursive predicate. The invariant was nearly complete, but the proofs could not be finished in time during the competition.

\team{sw} submitted a direct implementation of \autoref{lst:smaller-nearest-algo}'s algorithm in VerCors. They represented stacks using VerCors's mathematical sequences (an approach that worked well because these are well supported by the background prover). They wrote \emph{pop} and \emph{peek} functions to manipulate sequences as stacks; and equipped them with contracts so that they could be used inside the main algorithm (for lack of time, they did not provide an implementation of \emph{pop}).
They progressed quite far in the verification activities, but were not able to complete the proof of part~A's third task during the competition.
While VerCors has no specific limitations that would have prevented them from completing the proof given more time (the invariant required for verifying the third task is quite involved), the team's participants remarked that invariant generation features would have been useful to speed up their work.

\team{yvette} and \team{heja} implemented in C the algorithm of part~A, and annotated it using ACSL comments. While \team{yvette} implemented the algorithm as described in the challenge, \team{heja} wrote a simpler, quadratic-time algorithm, which searches for the nearest smaller value to the left by iterating in reverse over the input sequence (that is, by literally following the definition of left neighbor). Both teams managed to complete the first verification task using Frama-C's WP plugin, but they could not complete the other tasks in the time during the competition. In particular,
difficulties with formalizing \emph{aliasing} among data structures used by the algorithm and proving absence of side effects---a result of C's low-level memory model---slowed the teams down and hindered further progress.

\team{eindhoven} managed to verify part A entirely using the mCRL2 model checker, but had to introduce restrictions on the cardinality of the input values due to the nature of their verification tool. Their proofs assume lists of up to six values; and each value ranges over four possible values. With these restrictions, they managed to complete all three verification tasks in less than an hour. In particular, the third verification task did not cause any particular trouble as model checking does not need manually-provided invariants.

\subsection{Postmortem Evaluation of the Challenge}
\label{ch2:post-eval}
\fi

We presented challenge~2 under the assumption that its part~A was somewhat easier
and more widely feasible than part~B.
The fact that most teams worked on part~A may seem to confirm 
our assumption about its relatively lower difficulty.\footnote{After the competition, \team{vercors} explained that they missed our hint that part~A was simpler, and chose part~B only because it looked like a different kind of challenge (as opposed to part~A, which they felt was similar in kind to challenge 1's part~A).
  In the heat of the competition, participants may miss details of the challenges that may have helped them;
  this is another factor that should be considered when designing a challenge.}
At the same time, one out of only two teams who submitted a complete and correct solution to challenge~2 tackled part~B.
This may just be survival bias  but another plausible explanation is that the difficulties of the two parts are not so different (even though part~B looks more involved).

Indeed, part~A revealed some difficulties that were not obvious when we designed it. First, the algorithm in \autoref{lst:smaller-nearest-algo} follows an imperative style, and hence, it is not obvious how to encode it using functional style; various teams introduced subtle mistakes while trying to do so. Part~B is easier in this respect,
as the Cartesian tree construction algorithm consists of a simple iteration over the input, which manipulates data that can all be encoded indifferently using sequences, arrays, or lists.
Part~A, in contrast, requires a stack data structure with its operations.
In the end, 
what really makes part~B harder than part~A is probably its third, optional, verification task \emph{traversal}.
Specifying it is not overly complicated, but proving it requires a complex ``big'' invariant, which was understandably not easy to devise in the limited time available during the competition.

\section{Challenge 3: Sparse Matrix Multiplication}
\label{sec:challenge3}

The third challenge targeted the \emph{parallelization} of a basic algorithm to multiply \emph{sparse} matrices
(where most values are zero).

\subsection{Challenge Description}
\label{ch3:description}

We represent \emph{sparse matrices} using the coordinate list (COO) format. In this format, nonzero values of a matrix are
stored in a sequence of triplets, each containing row, column, and corresponding value. The sequence is sorted, first by row index and then by column index, for faster lookup. For example, the matrix:
\[
  \left( {\begin{array}{cccc}
   0 & 0 & 1 & 0 \\
   5 & 8 & 0 & 0 \\
   0 & 0 & 0 & 0 \\
   0 & 3 & 0 & 0 \\
  \end{array} } \right)
\]
is encoded into the following sequence (using row and column indexes that start from 1):
\[ (1, 3, 1)\ (2, 1, 5)\ (2, 2, 8)\ (4, 2, 3) \]

In this challenge, we consider an algorithm that computes the multiplication of a vector of values (encoded as a sequence) with a sparse matrix. It iterates over the values present inside the matrix, multiplies each of them by the appropriate element in the input vector, and stores the result at the appropriate index in the output vector. \autoref{lst:mult-algo} presents the algorithm in pseudo-code. 

\begin{figure}[tb]
\begin{lstlisting}
 y := (0, ..., 0)
 for every element (r, c, v) in m:
    y (c) := y (c) + x (r) * v
\end{lstlisting}
  \caption{Algorithm to multiply an input vector \lstinline|x| with a sparse matrix \lstinline|m| and store the result in the output vector \lstinline|y|.
  Input matrix \lstinline|m| is represented in the COO format as a list of triplets.}
  \label{lst:mult-algo}
\end{figure}

To solve challenge~3, we asked participants to carry out the following tasks:

\begin{description}
 \item[\textbf{Implementation}] tasks.
\begin{enumerate}
 \item Implement the algorithm to multiply a vector $x$ with a sparse matrix $m$.
 \item We want to execute this algorithm in parallel, so that each computation is done by a different process, \linebreak thread, or task. Add the necessary synchronization steps in your sequential program, using the synchronisation feature of your choice (lock, atomic block, \ldots).
 
 You can choose how to allocate work to processes. For example:
\begin{itemize}
 \item each process computes exactly one iteration of the for loop;
 \item there is a fixed number of processes, each taking an equal share of 
the total number of for loop iterations;
 \item work is assigned to processes dynamically (for example using a work stealing algorithm).
\end{itemize}
\end{enumerate}

 \item[\textbf{Verification}] tasks.
\begin{enumerate}
 \item Verify that the sequential multiplication algorithm indeed performs standard matrix multiplication (that is, it computes the output vector $y$ with values $y_i = \sum_{k} x_k \times m_{k,i}$).

 \item Verify that the concurrent algorithm does not exhibit concurrency issues (data races, deadlocks, \ldots).
 \item Verify that the concurrent algorithm still performs the same computation as the sequential algorithm.
If time permits, you can also experiment with different work allocation policies and verify that they all behave correctly.
\end{enumerate}
\end{description}

\iflong
\subsection{Designing the Challenge}
\label{ch3:design}

Since we designed challenge~3 last, after refining the description of the other two challenges, we ended up with several desiderata for it.

We wanted challenge~3 to target a concurrent algorithm,
but in a way that the challenge remained feasible, at least partly, also by participants using tools without explicit support for concurrency.
Expecting widely different degrees of support for concurrency,
we looked for a problem that was not completely trivial for teams using model-checking tools, which typically have built-in notions of concurrent synchronization and are fully automated.
Finally, true to the household style of VerifyThis competitions,
we wanted a problem that also involved behavioral (safety) input/output properties, as opposed to only pure concurrency properties like absence of deadlock and data races.

With the content of challenge~2 still fresh in our minds,
we first briefly considered some parallel algorithms to construct Cartesian trees.
It was soon clear that these would have added more complexity on top of an already challenging problem, and would have strongly penalized teams who found, for whatever reason, the Cartesian tree topic unpalatable.

Since even a modicum of concurrency significantly complicates the behavior of an algorithm, we decided to start from a sequential algorithm that was straightforward to understand.
The first candidate was a trivial problem where different processes increment a shared variable.
In a sequential setting, when processes execute one after another, the behavior is very simple to reason about.
But if the processes are allowed to interleave (that is, they run in parallel), some increments may be lost due to interference.
The issue with this problem is that verifying its concurrent behavior requires reasoning about the behavior of a program with races, but most verification frameworks for concurrent programs are geared towards proving the \emph{absence} of race conditions---so that
the input/output behavior of the overall program is independent of an execution schedule.
Therefore, being able to reason about the behavior of a program with races seemed
unsuitable.

Continuing from this observation in our search for a problem,
we ended up considering the matrix multiplication problem.
To avoid requiring to represent bidimensional data structures
we decided to target \emph{sparse} matrices, whose nonzero elements
can be encoded with a list of triples.

The standard sequential algorithm to multiply matrices is neither overly hard nor trivial, therefore it seemed a good basis for the challenge.
Parallelizing it is not \emph{conceptually} difficult;
however, we decided to give plenty of freedom in how computations are assigned to concurrent units (processes, threads, or tasks) both to accommodate different tools and to allow participants using tools with advanced support for concurrency to come up with sophisticated parallelization strategies and proofs.

As a final sanity check, we worked out a solution of this challenge using the model checker Spin.
ProMeLa---Spin's modeling language---offers primitives to model nondeterministic processes and to synchronize them, but also has limitations such
as support of only simple data types.
These features---typical of finite-state verification tools---made solving
challenge~3 possible in a reasonable amount of time
but certainly non-trivial.
In particular, we had to encode parts of the state model in C,
and then to
finesse the link between these foreign-code parts and the core ProMeLa model
so that the size of the whole state-space would not blow up during model checking.

Finally, we revised the description of challenge~3 to make sure that it was not biased towards any particular approach to modeling or reasoning about concurrency,
and that its sequential part was clearly accessible as a separate verification problem.
\fi

\subsection{Submitted Solutions}
\label{ch3:results}

\iflong
\subsubsection*{Overall Results} \fi

No teams solved challenge~3 completely. Six teams, out of the 12 teams\footnote{That is, one team skipped the last session.} that took part in VerifyThis's third and final session, attempted the verification of the sequential algorithm only---usually because their tools had little or no support for concurrency; out of these six teams, one completed verification task~1. Another six teams introduced concurrency in their implementation and tried to verify the absence of concurrency issues (verification task~2). Some of these teams used tools with built-in support for the verification of concurrent algorithms, while others added concurrency to their mostly sequential tools via custom libraries. Three teams out of the six that tackled task~2 completed the verification task in time during the competition; all of them happened to use a tool with built-in support for concurrency.
Finally, five teams attempted verification task~3 (proving that the sequential and concurrent algorithms compute the same output). Two of them achieved substantial progress on the proofs of task~3: \team{eindhoven} used a model checker with native support for concurrency;
\team{refiners} used Isabelle---a tool without built-in support for \linebreak concurrency---and hence modeled the concurrent implementation as a sequential algorithm that goes
over the sparse matrix's elements in nondeterministic order.

\iflong
\subsubsection*{Detailed Results}

Only teams using tools without support for concurrency attempted the verification of the sequential algorithm. Their implementations were close to the simple algorithm in \autoref{lst:mult-algo}---in some cases using recursion instead of looping. Verification task~1 (prove the correctness of the sequential matrix multiplication algorithm) required to specify the expected output given by ``standard matrix multiplication''. The approaches to expressing this property were quite varied.

\team{mergesort}, using Why3, defined a sparse matrix as a record containing two fields: a regular field (representing the sparse matrix in COO format) and a ghost field, representing the same matrix as a standard bidimensional array (with explicit zero values). A type invariant links together the two fields so that they represent the same matrix. The type invariant does not require uniqueness of indexes in the COO representation; if the element at a certain row and column appears more than once in the input sequence, its value in the ``standard'' matrix is taken to be the \emph{sum} of values in all such occurrences.
\team{yvette}, using Frama-C, introduced the ``standard'' matrix as an additional parameter of the multiplication function. The predicate linking the two representations was straightforward, stating that all elements in the COO representation are in the matrix, and that any elements of the matrix not in COO representation are zero. Uniqueness of indexes in the input sequence follows by assuming that they are ordered.
\team{kiv} followed a different approach to ensure uniqueness of indexes: they represented the input sparse matrix by means of a map instead of a list. For ``standard'' matrices, they went for arrays of arrays, as KIV does not have support for multi-dimensional arrays.
\team{mergesort}, \team{yvette} and \team{kiv}
achieved good results in producing accurate specifications,
but they did not have enough time left to complete the verification task during the competition.

Several teams who used tools without built-in support for concurrency
still managed to model concurrent behavior indirectly by making the order in which input elements are processed nondeterministic.
\team{viper} defined axiomatically a summation function over sets,
and used it to specify progress:
at any time during the computation,
a set variable stores the elements of the input that have been processed so far;
the current value of the output is thus the sum involving all the matrix elements in that set.
This specification style has the advantage of being independent of the order in which input elements are processed, and thus it encompasses both the sequential and the concurrent algorithms. By the end of the competition, \team{viper} got close to completing the corresponding correctness proofs.

Following a somewhat similar idea, \team{coinductive} im\-ple\-ment\-ed two versions of the multiplication algorithm: one operating directly on the COO list, and the other on a binary tree. The tree defines a specific order in which elements are processed and combined to get the final result, corresponding to different execution schedules.
Then, \team{coinductive} proved a lemma stating that both versions of the algorithm compute the same output---with some unproved assumptions about the associativity of vector addition.

\team{refiners} used Isabelle's refinement framework to prove that the sequential algorithm for multiplication of sparse matrices (\autoref{lst:mult-algo}) was a refinement of the ``standard'' multiplication algorithm on regular matrices.
Then, to lift their proofs to the concurrent setting, they modified the sequential algorithm so that it inputs a multiset instead of a list.
Since the order in which a multiset's elements are processed is nondeterministic,
the modified algorithm models every possible concurrent execution.
They also started modeling a work assignment algorithm (as an implementation of a folding scheme over the multisets),
but they did not completely finish the proofs of this more advanced part.

In preparation for their participation in VerifyThis, \linebreak
\team{bashers} developed a library for verifying concurrent programs in Isabelle, which they could deploy to solve challenge~3.
The library supported locking individual elements of an array.
Unfortunately, this granularity of locking turned out to be too fine grained for challenge~3, and they struggled to adapt it to model the algorithm of challenge~3 in a way that worked well for verification.

Among the tools used in VerifyThis 2019, three had built-in support for concurrency:
VerCors (using separation logic), Iris (a framework for higher-order concurrent separation logic in Coq), and the model checker mCRL2. The four teams using these tools---\team{vercors}, \team{sw}, \team{jm}, and \team{eindhoven}---managed to encode the concurrent algorithm, and to verify, possibly under simplifying assumptions, that it does not exhibit concurrency issues (verification task 2).

\team{jm}, using Coq's Iris, verified the safety of a single arbitrary iteration of the concurrent loop in \autoref{lst:mult-algo}. They encoded the concurrent algorithm using a deeply embedded toy language named LambdaRust, which features compare-and-set instructions as synchronization primitives. They ran out of time trying to extend the proof to all iterations of the loop.

Both teams using VerCors followed the same strategy of implementing the concurrent multiplication algorithm using parallel loops and an atomic block around the output update (the loop's body) to avoid interference. Thanks to VerCors's features, they had no major difficulties verifying that the code does not exhibit concurrency issues. Progress in task~3---verifying the functional behavior of the algorithm---was more limited.
A major stumbling block was that VerCors does not have support for summation over collections of elements; introducing and specifying this feature (required for task~3) was quite time-consuming.
\team{vercors} set up the algorithm's functional specification by introducing a summation function without specifying it fully---which limited the extent of what could be proved.
Their specification used ghost variables to encode the input's matrix in ``regular'' form, as well as a mapping between this form and the COO input sequence in sparse form.
The mapping explicitly defined an element in the COO sequence for every nonzero element of the full matrix, so that no existential quantification is needed.

\team{eindhoven} was the only team that completed verification of task~3, albeit with the usual simplifying assumptions (on input size and on the number of processes) that are required by the finite-state nature of model checkers.
They explicitly built the ``standard'' matrix equivalent of the input sparse matrix, and verified that the output was the expected result for all possible finitely many interleavings (which are exhaustively explored by the model checker).
If they had had more time,
they remarked that they would have tried to \emph{validate} their model:
the proofs assert the equivalence of two implementations,
but it would be best to perform a sanity check that they work as expected.

\subsection{Postmortem Evaluation of the Challenge}
\label{ch3:post-eval}
\fi

Regardless of whether their verification tools supported concurrency,
all teams had plenty of work to do in challenge~3.
We wanted a challenge that was approachable by everybody,
and it seems that challenge~3 achieved this goal.

On the other hand, the challenge turned out to be more time-consuming than we anticipated.
The sequential and the concurrent part \emph{alone} were enough to fill all 90 minutes of the competition session, and no team could complete the whole challenge.\footnote{Using a model checker, \team{eindhoven} covered all verification tasks but relied on simplifying assumptions on input size and number of processes.}
When we designed the challenge, we did not realize how time-consuming it would be.

The multiplication algorithm is conceptually simple,
but verifying it requires to fill in a lot of details, such as associativity and commutativity properties of arithmetic operations, that are not central to the algorithm's behavior but are necessary to complete the proofs.
In most cases, it was these details that prevented participants from completing the challenge.
Another feature that is often missing from verification tools but was required for challenge~3 is the ability of expressing sums over sets and sequences; while this can always be specified and verified, doing so takes time and distracts from the main goal of the challenge.

In all, verification challenges involving concurrency are not only harder to verify but also to design!
There are so many widely different concurrency models and verification frameworks
that calibrating a challenge so that it suits most of them is itself a challenge.
A possible suggestion to come up with concurrency challenges in the future is to write problems with different parts that are suitable for different verification approaches.
This strategy worked to ensure that tools without support for concurrency still had work to do in this challenge, and it may be generalizable to encompass different styles of concurrent programming and reasoning.

\section{Discussion}
\label{sec:discussion}

We organize the discussion around \iflong four \fi themes.
\autoref{sec:revised} outlines how teams revised their solutions for publication in the months after the competition.
\iflong \autoref{sec:tools} points out some tool features that emerged during VerifyThis 2019,
and briefly discusses how they relate to open challenges in verification technology. \fi
\autoref{sec:difficulty} analyzes the features of the verification challenges offered over the years, and how they affect the teams' success rate.
\autoref{sec:lessons} mentions some lessons we learned during this year's VerifyThis, which we would like to pass on to future organizers.

\subsection{Revised Solutions}
\label{sec:revised}

A couple of weeks after VerifyThis was over, we contacted all participants again, asking them permission to publish their solutions online.
Teams who consented had the choice of either publishing the solutions they submitted during the competition or supplying revised solutions---which they could prepare with substantially more time and the benefit of hindsight.
Nine teams submitted revised solutions---from light revisions to significant extensions.
Among the former, \linebreak \team{jm} and \team{openjml} cleaned up their code, added comments, and improved a few aspects of the implementation or specification to make them more readable.
\team{yvette} thoroughly revised their solutions and filled in missing parts of specification and proofs, so as to complete parts~A of challenges~1 and~2, and the sequential part of challenge~3.
\team{kiv} and \team{viper} went further, as they also completed the concurrent part of challenge~3.
So did \team{vercors}, \team{sw},\footnote{\team{vercors} and \team{sw} worked together to prepare one revised solution that merged both teams' independent work during the competition.} and \team{refiners}
who also provided partial solutions for part~B of challenge~2.
\team{mergesort} submitted extensively revised solutions, including the only complete solution to challenge~2's part~B---relying on a Coq proof of task \emph{traversal}\footnote{The proof obligation was generated automatically by Why3, but the Coq proof steps were supplied manually.}---and the sequential part of challenge~3.

The process of revising and extending solutions \emph{after} the competition is very different from that of developing them from scratch \emph{during} it.
With virtually unlimited time at their disposal, and the freedom to explore different approaches even if they may not pan out in the end, every team could in principle come up with a complete and correct solution.
At the same time, comparing the post-competition work of different teams is not very meaningful since some may simply not have additional time to devote to improving solutions---after all, we made it clear that revising solutions was something entirely optional that they did not commit to when they signed up for the competition.

\iflong
\subsection{Used Tools and Features}
\label{sec:tools}

Undeniably, SMT solvers have been a boon to verification technology, but some of their limitations may be a source of frustration even for users with lots of experience.
\team{openjml} reported the well-known problem of unresponsive proof attempts:
when a proof attempt is taking a long time, the user has to decide whether to abort it or to wait longer---hoping to get some counterexample information that may help debug the failed verification attempt.

Nowadays, SMT solvers are not limited to so-called auto-active~\cite{auto-active} tools such as OpenJML but also boost the level of automation of interactive provers.
The Isabelle proof assistant, for instance, extensively uses its Sledgehammer feature to automate the most routine proof steps (which make up a very large percentage of a typical proof).
Once an SMT solver manages to close the proof of some branch, Isabelle performs \emph{proof reconstruction} in order to generate a verifiable certificate of the SMT proof so that it can be soundly integrated into the overall Isabelle proof. \team{refiners}, using the Isabelle theorem prover, found the proof reconstruction step to be very time consuming in some cases, which even prevented them from completely closing the proof of some steps in time during the competition---even when they were confident in the SMT solver's results.

Using a high-level programming language (for example, one with an expressive type system) lets users focus on verifying complex behavioral properties on top of basic correctness properties (such as memory safety)---which are guaranteed by the language's semantics.
In contrast, when using a lower-level programming language, a large fraction of verification effort has to be devoted to establishing such basic properties---a time-consuming activity which may stifle progress towards more advanced verification goals.
We witnessed this phenomenon during the competition, when teams using C or other relatively low-level languages spent the majority of the time at their disposal verifying memory separation properties, while their colleagues using high-level languages could just assume them and jump right to the key input/output properties required by the problem statement.

VerifyThis has included a challenge involving \emph{concurrency} since its 2015 edition.
While it is probably still true that the majority of program verification tools focus on sequential correctness,
features to reason about concurrency and parallelism are increasingly available.
Despite this undeniable progress, verifying concurrent program remains for\-mi\-da\-bly difficult, and even participants using a tool expressly designed to reason about concurrency (such as VerCors and Viper, both supporting a permission logic) spent a considerable amount of time choosing what kind of synchronization primitive to use and how to model them in a formal way.

Model checkers are in a league of their own, since they are more similar to the tools used in fully-automated verification competitions such as SV-COMP than to the auto-active or interactive tools that most participants to VerifyThis prefer.
A model checker performs proofs completely automatically (crucially, it does not
require users to supply invariants) and is very effective at finding
errors when they exist.
Modeling concurrency is also typically straightforward, since the programming language of a model checker is typically built around a transparent model of parallel processes that communicate through shared memory, message passing, or both.
On the flip side, model checkers can only explore finite state spaces, and hence cannot normally perform verification of algorithms on unbounded data structures.
While these differences complicate judging teams using model checkers on par with others, we believe that having teams using model checkers taking part in VerifyThis adds depth to the competition, and strengthens the connections with other verification competitions while emphasizing its own peculiarity and focus.

\begin{table}[tb]
  \centering
  \scriptsize
  \setlength{\tabcolsep}{1.5pt}
  \begin{tabular}{l *{7}{r} |r}
 & \multicolumn{7}{c}{\textsc{VerifyThis competition}} \\
\cmidrule(lr){2-8}
 & \multicolumn{1}{c}{2011} & \multicolumn{1}{c}{2012} & \multicolumn{1}{c}{2015} & \multicolumn{1}{c}{2016} & \multicolumn{1}{c}{2017} & \multicolumn{1}{c}{2018} & \multicolumn{1}{c}{2019} & \multicolumn{1}{c}{\textsc{by tool}} \\
\midrule
AProVe & 1 & 0 & 0 & 0 & 0 & 0 & 0 & 1 \\
AutoProof & 0 & 0 & 1 & 0 & 0 & 0 & 0 & 1 \\
CBMC & 0 & 0 & 1 & 0 & 0 & 0 & 0 & 1 \\
CIVL & 0 & 0 & 0 & 1 & 1 & 1 & 0 & 3 \\
Coq & 0 & 0 & 0 & 0 & 0 & 1 & 2 & 3 \\
Dafny & 1 & 2 & 3 & 5 & 1 & 1 & 0 & 13 \\
ESC/Java2 & 0 & 1 & 0 & 0 & 0 & 0 & 0 & 1 \\
F* & 0 & 0 & 1 & 0 & 0 & 0 & 0 & 1 \\
Frama-C & 0 & 0 & 1 & 0 & 1 & 1 & 2 & 5 \\
GNATProve & 0 & 1 & 0 & 0 & 0 & 0 & 0 & 1 \\
Isabelle & 0 & 0 & 0 & 0 & 0 & 1 & 2 & 3 \\
jStar & 1 & 0 & 0 & 0 & 0 & 0 & 0 & 1 \\
KeY & 1 & 1 & 1 & 1 & 2 & 1 & 0 & 7 \\
KIV & 1 & 1 & 1 & 1 & 1 & 1 & 1 & 7 \\
mCRL2 & 0 & 0 & 1 & 1 & 0 & 0 & 1 & 3 \\
MoCHi & 0 & 0 & 1 & 0 & 0 & 0 & 0 & 1 \\
OpenJML & 0 & 0 & 0 & 0 & 0 & 0 & 1 & 1 \\
PAT & 0 & 1 & 0 & 0 & 0 & 0 & 0 & 1 \\
VCC & 0 & 1 & 0 & 0 & 0 & 0 & 0 & 1 \\
VerCors & 0 & 0 & 1 & 1 & 1 & 2 & 2 & 7 \\
VeriFast & 0 & 1 & 1 & 1 & 0 & 1 & 0 & 4 \\
Viper & 0 & 0 & 0 & 1 & 0 & 1 & 1 & 3 \\
Why3 & 1 & 2 & 1 & 2 & 3 & 1 & 1 & 11 \\
\midrule
\multicolumn{1}{c}{\textsc{by competition}} & 6 & 9 & 12 & 9 & 7 & 11 & 9
\end{tabular}
   \caption{Verification tools used in each VerifyThis competition. Number $n$ in row $t$ and column $y$ means that $n$ teams used tool $t$ during VerifyThis $y$. The rightmost column (\textsc{by tool}) reports the total number of teams that used each tool; and the bottom row (\textsc{by competition}) reports how many different tools were used in each competition.}
  \label{tab:tools-history}
\end{table}

\paragraph{Trends in tool usage.}

\autoref{tab:tools-history} updates the data about tools used at VerifyThis~\cite{vt-retrospective}.
A total of 23 tools were used in VerifyThis competitions to date.
A team has used one single tool in all cases except for a single-person team that used two tools (Dafny and KeY) during VerifyThis 2018.
Teams winning the ``best overall'' award used one of three tools: Isabelle, VeriFast, and Why3---each tool used by two awardees.
The tools used by winners of ``best student team'' include Dafny, KIV, mCRL2 (one winning team each), VerCors (two winning teams), and Why3 (five winning teams).
Several tools were singled out for having a ``distinguished feature'' that deserved an award because it was apt to tackle some verification challenges: CIVL, GNATProve, Isabelle, KIV, MoCHi, VerCors, Viper, and Why3.

\fi

\begin{table*}[ht]
\centering
\scriptsize
\setlength{\tabcolsep}{2pt}
\begin{tabular}{llrrlllllrr}
  \toprule
{\textsc{problem}} & {\textsc{competition}} & {\textsc{order}} & {\textsc{time}} & {\textsc{sequential}} & {\textsc{input}} & {\textsc{output}} & {\textsc{algorithm}} & {\textsc{mutable}} & {\textsc{partial}} & {\textsc{complete}} \\ 
  \midrule
Maximum by elimination & VT11 & 1 & 60 & sequential & array & simple & iterative & immutable & 83 & 67 \\ 
  Tree maximum & VT11 & 2 & 90 & sequential & tree & simple & outlined & immutable & 100 & 17 \\ 
  Find duplets in array & VT11 & 3 & 90 & sequential & array & simple & find & immutable & 83 & 50 \\ 
  Longest common prefix & VT12 & 1 & 45 & sequential & array & simple & iterative & immutable & 100 & 73 \\ 
  Prefix sum & VT12 & 2 & 90 & sequential & array & complex & outlined & immutable & 73 & 9 \\ 
  Delete min node in binary search tree & VT12 & 3 & 90 & sequential & tree & same & recursive & mutable & 18 & 0 \\ 
  Relaxed prefix & VT15 & 1 & 60 & sequential & array & same & iterative & immutable & 79 & 7 \\ 
  Parallel GCD by subtraction & VT15 & 2 & 60 & concurrent & scalar & same & iterative & mutable & 79 & 14 \\ 
  Doubly linked lists & VT15 & 3 & 90 & sequential & linked list & same & outlined & mutable & 71 & 7 \\ 
  Matrix multiplication & VT16 & 1 & 90 & sequential & matrix & same & iterative & immutable & 86 & 43 \\ 
  Binary tree traversal & VT16 & 2 & 90 & sequential & tree & simple & iterative & mutable & 79 & 7 \\ 
  Static tree barriers & VT16 & 3 & 90 & concurrent & tree & simple & iterative & mutable & 79 & 7 \\ 
  Pair insertion sort & VT17 & 1 & 90 & sequential & array & same & iterative & mutable & 100 & 10 \\ 
  Odd-even transposition sort & VT17 & 3 & 90 & concurrent & array & same & iterative & mutable & 60 & 0 \\ 
  Tree buffer & VT17 & 4 & 90 & sequential & tree & same & recursive & immutable & 40 & 20 \\ 
  Gap buffer & VT18 & 1 & 60 & sequential & array & same & iterative & mutable & 91 & 36 \\ 
  Count colored tiles & VT18 & 2 & 90 & sequential & array & same & recursive & immutable & 55 & 18 \\ 
  Array-based queue lock & VT18 & 3 & 90 & concurrent & array & simple & iterative & mutable & 27 & 9 \\ 
  Monotonic segments and GCG sort & VT19 & 1 & 90 & sequential & array & same & iterative & immutable & 85 & 15 \\ 
  Cartesian trees & VT19 & 2 & 90 & sequential & array & complex & iterative & immutable & 69 & 15 \\ 
  Sparse matrix multiplication & VT19 & 3 & 90 & concurrent & matrix & same & iterative & immutable & 85 & 0 \\ 
   \bottomrule
\end{tabular}
\caption{For each challenge \textsc{problem} used at VerifyThis: the \textsc{competition} when it was used; the \textsc{order} in which it appeared; how much \textsc{time} (in minutes) was given to participants to solve it; whether the main algorithm is \textsc{sequential} or concurrent; the main \textsc{input} data type; whether the \textsc{output} data type is of the \textit{same} kind as the input, \textit{simple}r, or more \textit{complex}; when the \textsc{algorithm} was given in pseudo-code, whether it was \textit{iterative} or \textit{recursive} (if it was not given, whether it was \textit{outlined} or participants had to \textit{find} it based on the requirements); whether the input is \textsc{mutable} or immutable; and the percentages of participating teams that were able to submit a \text{partial} or \textsc{complete} solution.} 
\label{tab:problems-classification}
\end{table*}

\subsection{What Makes a Challenge Difficult?}
\label{sec:difficulty}

We used various criteria to classify the 21 challenges used at VerifyThis to date---three in each edition excluding VerifyThis 2014, which was run a bit differently among participants to a Dagstuhl Seminar.
We classified each challenge according to which VerifyThis \pred{competition} it belongs to, whether it appeared first, second, or third in \pred{order} of competition, how much \pred{time} was given to solve it, whether it targets a \pred{sequential} or concurrent algorithm, what kind of \pred{input} data it processes (array, tree, linked list, and so on), whether the main algorithm's \pred{output} involves the same kind of data structure as the input, whether the challenge's main \pred{algorithm} is iterative or recursive (or if the algorithm is only outlined), and whether the input data structure is mutable or immutable.
For each challenge, we also record what percentage of participating teams managed to submit a partial or complete correct solution.
\autoref{tab:problems-classification} shows the results of this classification.

To help us understand which factors affect the complexity of a verification problem,
we fit a linear regression model (with normal error function) that uses \pred{competition}, \pred{order}, \pred{time}, \pred{sequential}, \pred{input}, \pred{output}, \pred{algorithm}, and \pred{mutable} as predictors, and the percentage of \pred{complete} solutions as outcome.\footnote{We could also perform a similar analysis using the percentage of \pred{partial} solutions as outcome. However, what counts as ``partially correct'' is a matter of degree and depends on a more subjective judgment---which would risk making the analysis invalid.}
Using standard practices~\cite{GelmanHill-linearmodels}, categorical predictors that can take $n$ different values are encoded as $n-1$ binary indicator variables---each selecting a possible discrete value for the predictor.
Fitting a linear regression model provides, for each predictor, a regression coefficient estimate and a standard error of the estimate;
the value of the predictor has a definite effect on the outcome if the corresponding coefficient estimate differs from zero by at least two standard errors.

Our analysis suggests that the competition challenges were somewhat simpler in the early editions compared to the recent editions (starting from VerifyThis~2015): the coefficients for indicator variables related to predictor \pred{competition} for the years 2015--2017 and 2019 are clearly negative,
indicating that belonging to one of these editions tends to decrease the number of correct solutions.
Similarly, the later a challenge problem appears in a competition the fewer teams manage to solve it correctly.
This is to be expected, as the first challenge is normally the simpler and more widely accessible one, and participants get tired as a competition stretches over several hours.

When a challenge's main algorithm is only outlined, or is given in pseudo-code but is recursive,
and when the input is a mutable data structure,
participants found it harder to complete a correct solution.
While the difficulty of dealing with mutable input is well known---and a key challenge of formal verification---different reasons may be behind the impact of dealing with naturally recursive algorithms.
One interpretation is that verification tools are still primarily \linebreak geared towards iterative algorithms;
a different, but related, interpretation is that VerifyThis organizers are better at gauging the complexity of verifying iterative algorithms,
as opposed to that of recursive algorithms that may be easy to present but hard to prove correct.

Sequential algorithms, as opposed to concurrent ones, are associated with harder problems.
Since the association is not very strong, it is possible that this is only a fluke of the analysis:
sequential algorithms are the vast majority (76\%) of challenges and thus span a wide range of difficulties;
the few challenges involving concurrent algorithms
have often been presented in a way that they offer a simpler, sequential variant
(see for example challenge~3 in \autoref{sec:challenge3})---which may be what most teams go for.

The input data structure also correlates with the ease of verification.
Unsurprisingly, when the algorithm's input is a scalar more teams are successful; but, somewhat unexpectedly, success increases also when the input is a linked list or a tree.
It is possible that the organizers are well aware of the difficulty of dealing with heap-allocated data structures,
and hence stick to relatively simple algorithms when using them in a verification challenge.
Another possibility is that linked lists and trees just featured a few times (compared to the ubiquitous arrays), and hence their impact is more of a statistical fluke.
Input in matrix form is associated with harder problems too; this is probably because most verification tools have no built-in matrix data types, and representing bidimensional data using arrays or lists is possible in principle but may be cumbersome.

\subsection{Lessons Learned for Future Competitions}
\label{sec:lessons}

Most verification tools are somewhat specialized in the kinds of properties and programs they mainly target;
such specialization normally comes with a cost in terms of less flexibility when tackling challenges outside their purview.
VerifyThis organizers try to select challenges that target different domains and properties, so that no participants will be exclusively advantaged.
However, this may also indicate that it may be interesting to see the participation of teams using \emph{different approaches}.
While only one team has used two different verification tools in the history of VerifyThis,
teams using verification frameworks that integrate different libraries and features effectively have at their disposal a variety of approaches.
For instance, \team{refiners} used a refinement library for Isabelle only in one challenge, whereas they stuck to Isabelle's mainstream features for the rest of the competition.
In order to promote eclectic approaches to verification, organizers of future events may introduce a new award category that rewards the teams that displayed the widest variety of approaches during the competition.

VerifyThis challenges are made publicly available after the competition every year, and several team members took part in more than one competition.
Therefore, the most competitive and ambitious teams are aware of the kinds of problems that will be presented, and may be better prepared to solve them in the limited time at their disposal.
We have evidence of at least one team that went one step further preparing for the competition this year: \team{bashers} created an Isabelle library to reason about concurrency, expecting a challenge of the same flavor as those given in recent years.
These observations may give new ideas to organizers of future events to design verification challenges that are interesting but also feasible.
For example, they could announce before the competition (in the call for participation) some topics that will appear in the verification challenges, or some program features that participants will be expected to deal with---but without mentioning specific algorithms or problems.
Researchers and practitioners interested in participating may then use this information to focus their preparation.

Following the recurring suggestions of previous organizers, we used a questionnaire to help compare solutions and judge them.
This was of great help and we hope future organizers can improve this practice even further.
While our questionnaire was primarily made of open questions and collected qualitative data, it may be interesting to complement it with \emph{quantitative} information about the challenges and the solutions---such as the size of specifications, implementation, and other tool-specific annotations.
Collecting such information consistently year after year could also pave the way for more insightful analyses of the trends in the evolution of verification technology as seen through the lens of verification competitions (perhaps along the lines of what we did in \autoref{sec:difficulty}).

We are always pleased to see how committed participants are, and how much effort they put into their work during and after the competition.
One sign of this commitment is that most teams (see \autoref{sec:revised} for details)
were available to substantially \emph{revise} their solutions during the weeks and months after the competition, so that we could publish a complete solution that shows the full extent of the capabilities of their tools.
It may be interesting to find ways to give more visibility to such additional work---for example, publishing post proceedings where teams can describe in detail their work and how it was perfected.
Since not all participants may be able to commit to such an extra amount of work, this may be organized only occasionally, and contributing to it should be on a voluntary basis.

\begin{acknowledgements}
We thank Amazon Web Services for sponsoring travel grants and monetary prizes;
Gidon Ernst, Nadia Polikarpova, and Stephen F.\ Siegel for submitting ideas for competition problems;
and Virgile Prevosto for accepting, on a short notice, our invitation to give a tutorial.
The extensive support by local organizers of ETAPS was fundamental for the practical success of VerifyThis.
Being part of TOOLympics further enhanced the interest and variety of the participants' experience,
and made VerifyThis more widely visible.
Finally, our biggest ``thank you'' goes to the participants for their enthusiastic participation, for their effort (often continued after the competition when preparing revised solutions), and for contributing to interesting discussions in an amiable atmosphere.
\end{acknowledgements}



\end{document}